\documentclass[%
 reprint,
superscriptaddress,
 amsmath,amssymb,
 aps,
prb,
floatfix]{revtex4-2}

\usepackage{soul,color}
\soulregister{\cite}{7}
\soulregister{\ref}{7}
\soulregister{\pageref}{7}

\usepackage{graphicx}
\usepackage{dcolumn}
\usepackage{bm}

\usepackage{physics}
\usepackage{float}

\begin{document}

\preprint{APS/123-QED}

\title{Polariton XY-simulators revisited}

\author{Junhui Cao}
\affiliation{%
 Abrikosov Center for Theoretical Physics, Moscow Center for Advanced Studies, Kulakova str. 20, Moscow, Russia
}%

\author{Denis Novokreschenov}
\affiliation{%
 Abrikosov Center for Theoretical Physics, Moscow Center for Advanced Studies, Kulakova str. 20, Moscow, Russia
}%

\author{Alexey Kavokin}
\email{a.kavokin@westlake.edu.cn}
\affiliation{%
 Abrikosov Center for Theoretical Physics, Moscow Center for Advanced Studies, Kulakova str. 20, Moscow, Russia
}%
\affiliation{Russian Quantum Center, Skolkovo, Moscow, 121205, Russia}
\affiliation{Department of Physics, St. Petersburg State University, University Embankment, 7/9, St. Petersburg, 199034, Russia}
\affiliation{School of Science, Westlake University, 18 Shilongshan Road, Hangzhou 310024, Zhejiang Province, China}


\date{\today}

\begin{abstract}
Arrays of bosonic condensates of exciton-polaritons have emerged as a promising platform for simulating classical XY models, capable of rapidly reaching phase-locked states that may be mapped to arrays of two-dimensional classical spins. However, it remains unclear whether these states genuinely minimize the corresponding XY Hamiltonian and how the convergence time scales with the system size. Here, we develop an analytical model revealing that an array of $N$ condensates possesses $N$ stable phase configurations. The system selectively amplifies a specific configuration dependent on the pump power: at low power, the state with the smallest eigenvalue of an effective XY Hamiltonian is favored, while at high power, the state with the largest eigenvalue prevails. At intermediate pump powers, the system visits all eigenstates of the Hamiltonian. Crucially, the formation rate for any of these phase-locked states remains on the order of 100 ps, independent of the size of the array, demonstrating the exceptional speed and scalability of polariton-based XY simulators.
\end{abstract}

\maketitle


\section{Introduction}

Exciton-polaritons (further also referred to as polaritons, for simplicity) are light-matter bosonic quasiparticles formed in semiconductor crystal structures by optical pumping or electronic injection\cite{amo2010exciton,byrnes2014exciton,graf2017electrical,keeling2011exciton,schneider2013electrically}. They combine properties of light and matter and exhibit ultrafast stimulated scattering dynamics\cite{chen2023unraveling,chestnov2025stimulated,tassone1999exciton,toffoletti2025coherent} that lead to formation of macroscopic coherent condensates of polaritons and the polariton lasing effect\cite{fraser2016physics,kavokin2003polariton,ramezani2016plasmon} associated to the phenomenon of non-equilibrium Bose-Einstein condensation.  Discoveries of superfluidity\cite{carusotto2004probing,amo2009superfluidity} and, recently, supersolidity of exciton-polaritons\cite{trypogeorgos2025emerging,nigro2025supersolidity}, works on polariton topological insulators, quantized vortices, half-vortices, solitons, half-solitons made polaritonics one of the most rapidly developing research fields of modern solid state physics\cite{klembt2018exciton,lagoudakis2008quantized,egorov2009bright,hivet2012half,sitnik2022spontaneous}. After the discovery of the most important fundamental effects of polaritonics in the first decade of the XXI century, much effort has been directed to finding a niche for their application in opto-electronics, communications, or computing\cite{zhao2023exciton,zhang2018photonic,ghosh2020quantum,kavokin2022polariton}. The potentiality of coherent polariton fluids for emulation of complex physical systems seems especially high, as polariton ensembles are very flexible, and they easily mimic a number of other physical systems from spin glasses to black holes\cite{shelykh2005spin,yang2022microcavity,solnyshkov2011black,gerace2012analog,askitopoulos2018all}. The use of arrays of exciton-polariton condensates as ultrafast analogue simulators for the solution of many-body interaction problems is one of the most tempting application proposals\cite{pistorius2020quantum,fowler2022efficient,PhysRevApplied.17.024063}.

Since the pioneering work of Berloff et al\cite{berloff2017realizing}, simulators based on arrays of Bose-Einstein condensates of exciton-polaritons distributed in a plane of a planar semiconductor microcavity attracted the attention of a multidisciplinary community of physicists and information scientists\cite{kim2017exciton,krupp2025quantum,chng2025quantum}. The idea behind a series of experimental papers devoted to polariton XY-simulations is to associate the phase of each polariton condensate to a two-dimensional vector (a classical spin) and study vector configurations corresponding to phase locked states of the condensate arrays. Experimentally, phase locking in arrays of polariton condensates has been achieved on a 100 ps or shorter time-scale which opens a tremendous opportunity to employ this effect for minimization of certain classical Hamiltonians. In particular, it was suggested that the phase-locked state corresponds to the ground state of an XY Hamiltonian based on scalar products of two-dimensional classical spins describing the phases of neighboring polariton condensates. The efficiency of polariton XY simulators was demonstrated on a number of specific configurations of the arrays of condensates and supported by simulations based on the generalized Gross-Pitaevskii equation coupled with rate equations for incoherent exciton reservoirs created by non-resonant laser pumping\cite{carusotto2013quantum,wouters2008spatial,nespolo2019generalized}. Essentially, the phase locking of polariton condensates was shown to be a consequence of the interplay of coherent and dissipative coupling effects, first described by Aleiner, Altshuler and Rubo\cite{PhysRevB.85.121301}.

Despite of the success of these proof-of-concept experiments several important questions remain to be answered, namely:
\begin{itemize}
    \item Is it true that any array of exciton polariton condensates at any value of pumping spontaneously finds a phase configuration that minimizes the corresponding XY Hamiltonian? If not, what the limitations are?
    \item How the time that the system spends to minimize the XY Hamiltonian scales with the number of condensates in the array? 
\end{itemize}   
In order to address these crucial questions we have developed an analytical model that describes formation of phase-locked modes in an array of polariton condensates at the early stages of its formation. In contrast to the previous theories based on the generalized Gross-Pitaesvkii equation coupled to the rate equations for incoherent exciton reservoirs, we describe the system in the framework of a tight-binding model, where coupling between individual exciton-polariton condensates is evaluated based on the overlap integrals of their wavefunctions. This approach allows us obtaining the full set of multi-condensate eigenmodes of the system, which constitutes an important difference with the Gross-Pitaevskii modeling that yields a single multi-condensate state corresponding to the dominating mode but fails to provide information on other modes in the spectrum. Solving rate equations for each of the eigenmodes of the system we analyze their formation dynamics as functions of the pump power.  

This analysis led us to several important conclusions. First, an array of $N$ polariton condensates can be characterized by $N$ stable phase configurations that are robust against acoustic phonon scattering and may be preserved on time scales many orders of magnitude longer than the lifetime of a single polariton. Second, each of these phase configurations can be characterized by a threshold pump power that corresponds to the onset of Bose-Einstein condensation. The threshold pump power is governed by the balance of the income rate of polaritons that come into the given mode from incoherent reservoirs and the radiative decay rate. The system is expected to first grow the mode with the lowest threshold pump power. The growth rate of the number of particles in a given phase configuration is governed by the difference between the rate of stimulated scattering of excitons from incoherent reservoirs to the polariton mode and the radiative decay rate. Although the decay rate is proportional to the total number of polaritons in a given mode, the stimulated scattering rate is governed by the number of polaritons located under the pump spots. If the pump spots are far enough from each other, the configuration characterized by the smallest part of the multi-condensate wavefunction spreading outside the pump spots will have the lowest threshold power. On the other hand, as the pumping intensity increases, the configuration with a largest overlap with pump spots will have the highest growth rate. Thus, in the limits of low and high pumping intensities, the build up of phase configurations corresponding to the lowest and largest eigenvalues of a matrix of overlap integrals of single-condensate wavefunctions is expected, respectively. At the intermediate pump powers, the system would be able to visit the intermediate eigen-states. It is very important to note that the matrix of overlap integrals which we analyze can be reduced to an effective XY Hamiltonian. This is cardinal for the use of arrays of polariton condensates as XY simulators.
 
Remarkably, in all cases the formation rate of a phase locked state remains extremely high, that is, of the order of a formation rate of an individual exciton-polariton condensate. This shows that polariton-based XY simulators operate on a time scale of 100 ps even for large numbers of condensates\cite{christmann2014oscillatory,berloff2017realizing}. This makes them competitive not only with classical computers but also with quantum simulators based on different material platforms.

\begin{figure}
    \centering
    \includegraphics[width=1\linewidth]{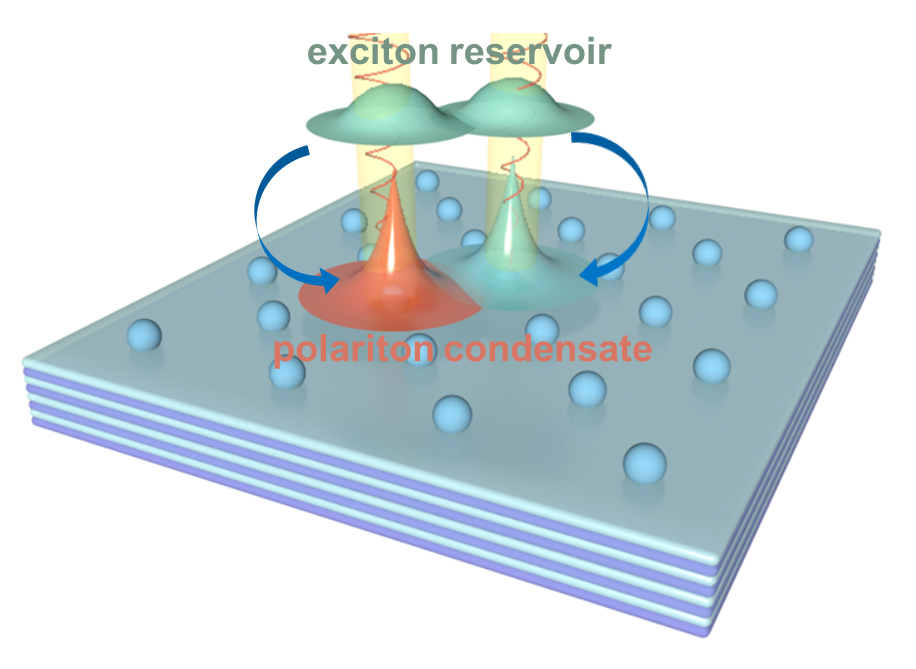}
    \caption{Schematic showing an array of coupled exciton-polariton condensates in a planar semiconductor microcavity. Non-resonant pump beams generate incoherent exciton reservoirs, which supply quasiparticles to the exciton-polariton modes. The phase locking of individual polariton condensates governed by coherent and dissipative coupling enables the system to function as an XY Hamiltonian simulator.}
    \label{fig_concept}
\end{figure}

\section{Theory and Method}
We consider an array of $N$ exciton-polariton condensates spread in a plane of a semiconductor microcavity (see the schematic in Figure 1). Each condensate is formed under the optical pump spot. We assume that the pumping is non-resonant, and it creates a cloud of incoherent excitons that, in turn, fed the exciton-polariton condensate by means of a stimulated scattering. Essentially, we consider the model system equivalent to one studied in Ref \cite{berloff2017realizing}. All pump beams are assumed to be of the same intensity.
We start with a trivial case of a single polariton condensate that has been extensively studied. Briefly revisiting the well-known expressions for the condensate wavefunction, we shall proceed to the consideration of an array of condensates.
In contrast to most previous studies\cite{nalitov2017spontaneous,chestnov2024nonadiabatic,nalitov2019optically}, we $do$ $not$ solve the generalized Gross-Pitaevskii equation for the condensate matrix. Instead, we find the eigenmodes of the empty quantum system built on the basis of $N$ known single-condensate wavefunctions spread in the cavity plane. This set of multicondensate eigenmodes can be found analytically by diagonalization of the matrix $D$ composed by the overlap integrals of our individual condensate wavefunctions. This algebraic procedure enables us to obtain a set of $N$ orthogonal multi-condensate states that are expected to demonstrate essentially independent time-dynamics at least in the low pumping regime where polariton-polariton interaction effects can be ignored. We describe each of these states by a rate equation including the pump and decay terms accounting for the stimulated scattering of excitons from the reservoirs and the radiative decay of exciton-polaritons, respectively. Analyzing these equations, we come to important conclusions on the pump-dependence of the build-up of populations of competing multi-condensate eigen modes of the system. 

\subsection{Model}
The stationary state of a single, freely propagating polariton condensate can be described by the Schrödinger equation for polaritons characterized by the effective mass $m$:
\begin{equation}
\left(-\frac{\hbar^2}{2m}\nabla^2-E\right)\Psi(\mathbf{r})=0,
\end{equation}
where $E=\hbar^2k_c^2/(2m)$ is the energy of the condensate and $k_c$ is the magnitude of its in-plane wavevector, generated from the repulsive interaction between polaritons and excitons. $k_c$ is proportional to the square root of the exciton density\cite{yagafarov2020mechanisms}. This equation is equivalent to the Helmholtz equation, whose fundamental solutions representing outgoing cylindrical waves are the Hankel functions of the first kind. Thus, the wavefunction for a condensate located at site $i$ with position $\mathbf{r}_i$ can be approximated by:
\begin{equation}
\Psi_i(\mathbf{r})=H_0^{(1)}(k_c|\mathbf{r}-\mathbf{r}_i|).
\end{equation}
This wavefunction exhibits a logarithmic divergence at the source point $\mathbf{r}_i$ and asymptotically behaves as an outgoing cylindrical wave:
\begin{equation}
\lim_{|\mathbf{r}-\mathbf{r}_i|\to\infty}H_0^{(1)}(k_c|\mathbf{r}-\mathbf{r}_i|)\approx\sqrt{\frac{2}{\pi k_c|\mathbf{r}-\mathbf{r}_i|}}e^{i(k_c|\mathbf{r}-\mathbf{r}_i|-\pi/4)}.
\end{equation}
To handle the non-square-integrability of this scattering state, we employ a box normalization $\int_L|\Psi_i(\mathbf{r})|^2d\mathbf{r}=1$ over a domain $L$ that encompasses the entire lattice.

It is important to note that while the approximation of a single condensate wavefunction by the Hankel function is a convenient approach used in many previous publications \cite{berloff2017realizing}, strictly speaking, it is exact only for a point-like source emitting polaritons in radial direction in the microcavity plane. Experimentally, the Gaussian pump spots are used most frequently, which is why a convolution of the Hankel function to the Gaussian function describing the spatial distribution of point-like sources would represent the single condensate wavefunction more accurately (see Appendix B). Furthermore, to account for the repulsion of polaritons from the exciton reservoir in the presence of pump and dissipation one would need to solve a generalized Gross-Pitaevskii equation for the condensate coupled to the diffusion equation for the reservoir numerically. In this study, we focus initially on the analytical model that implies the Hankel ansatz for the single-condensate wavefunction. This all enable us to obtain analytically the most important characteristics of the formation dynamics of multi-condensate modes such as the threshold pump power and the growth rate of the mode occupancy. Next, we discuss the impact of the finite size of the pump spots on the interference patterns formed by phase-locked condensates in the Appendix B.

In all cases, the key quantity governing the collective behavior of the lattice is the overlap integral between single-condensate wavefunctions centered at the different pump spots. Using the Hankel ansatz, we construct the overlap matrix $D$ whose elements are defined as:
\begin{equation}
D_{ij}=\langle\Psi_j|\Psi_i\rangle=\frac{\int_LH_0^{(2)}(k_c|\mathbf{r}-\mathbf{r}_j|)H_0^{(1)}(k_c|\mathbf{r}-\mathbf{r}_i|)d\mathbf{r}}{\int_LH_0^{(2)}(k_c|\mathbf{r}-\mathbf{r}_i|)H_0^{(1)}(k_c|\mathbf{r}-\mathbf{r}_i|)d\mathbf{r}}.
\label{Eq:OME}
\end{equation}
Here the denominator ensures box normalization by setting all diagonal terms $D_{ii}=1$.
We shall emphasize that the correspondence between the eigenvalues of the matrix $D$ and the occupation numbers of polariton modes is subject to a normalization condition (see Appendix A for the derivation of this relation). Here, as a simplest example, we have adapted $D_{ii}=1$ condition to calculate the elements of matrix $D$. A constant factor depending on the actual polariton condensate occupation number $N_{\mathrm{pop}}$ would appear before the matrix $D$ in a general case (hence the diagonal term would be $D_{ii}=N_{\mathrm{pop}}$). Importantly, this factor would not affect the eigenvectors of the matrix.
For a lattice with $N$ sites, $D$ is a matrix of $N\times N$ size. In the present model, we consider overlaps only between nearest neighbors, resulting in a sparse matrix structure. We note that the model can be easily extended to account for couplings of next-to-nearest and distant neighbors. 

Diagonalizing this overlap matrix yields a set of eigenstates $\varphi_\nu (\mathbf{r})$ and corresponding eigenvalues $\lambda_\nu$:
\begin{equation}
\varphi_\nu(\mathbf{r})=\sum_{i=1}^Nc_i^{(\nu)}\Psi_i(\mathbf{r}),
\end{equation}
where $\mathbf{c}^{(\nu)}=(c_1^{(\nu)},c_2^{(\nu)},\ldots,c_N^{(\nu)})^T$ is the $\nu$-th eigenvector. The eigenvalue $\lambda_\nu$ characterizes the interference of multiple polariton condensates within the mode $\nu$.

The diagonalization of the overlap matrix $D$ yields the set of phase-locked multi-condensate states which are orthogonal to each other. It is safe to assume that each of these states may remain stable on a long timescale, provided that the proper balance of pump and decay is satisfied. In contrast, any superposition of these eigenstates is likely to lose coherence and relax to one of the eigenstates on a time-scale of the coherence time of a bosonic condensate of exciton-polaritons, that is a few hundred of picoseconds. Note also that acoustic phonon induced relaxation between orthogonal eigenstates of the system is unlikely, as the energy splitting between these states is expected to be very small (on a sub-micro-eV energy scale)\cite{stanley1997resonant}. The probabilities of transitions between orthogonal states with emission or absorption of such low energy (and consequently long wavelength) acoustic phonons are expected to be very low, because of the parity selection rules\cite{Cao_2021}. To conclude, it is safe to assume that on a long time-scale the system would converge to one of N orthogonal eigenstates given by the diagonalization of the $N\times N$ matrix $D$ with N being the number of pump beams creating individual polariton condensates. In order to find out which one of these states would be chosen by the system, we derive the Boltzmann-type rate equations for each of the eigenstates of the system and compare the population dynamics of individual eigen-modes predicted by these equations.

\subsection{Driven-dissipative mode selection}
\begin{figure}
    \centering
    \includegraphics[width=1\linewidth]{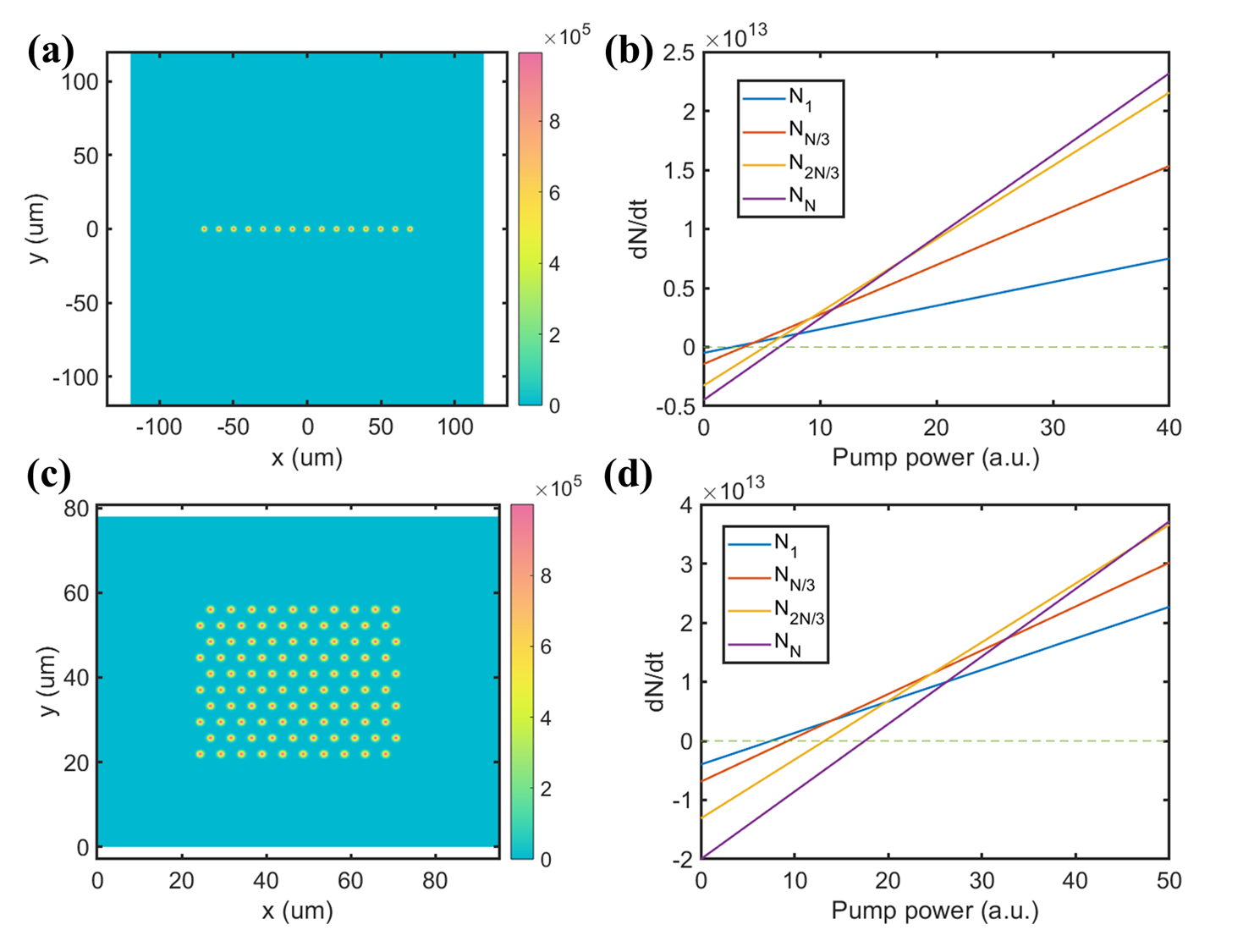}
    \caption{(a) Pump profile of a one-dimensional chain of exciton-polariton condensates. (b) The growth rate of the population of the polariton mode $N_\nu$ in a one-dimensional chain of polariton condensates with $N=15$. When pump power is low, the mode with a smallest eigenvalue $N_1$ exhibits the largest growth rate (blue curve). The increase of the pump power will eventually make the mode with the largest eigenvalue $N_N$ the fastest growing (purple curve). (c) Pump profile of a two-dimensional triangular zigzag polariton lattice. (d) The growth rate of the population of the polariton mode $N_\nu$ in the triangular lattice with $N=100$. }
    \label{fig_pump}
\end{figure}

In order to derive the rate equations for the eigenmodes of an array of $N$ coupled polariton condensates in the periodic lattice, we express the number of polaritons in a multi-condensate eigenstate $\varphi_\nu$ as a function of the reservoir density $n(r)$ accounting for the optical pumping rate $P$ and the dissipation rate in the reservoir $\gamma$, the stimulated scattering of polaritons from the reservoir to the condensate and the spontaneous radiative decay of polaritons from the condensate characterized by a rate $1/\tau$.

The rate equation for the population of a multi-condensate mode $N_\nu$ coupled to the rate equation for the coordinate-dependent density of excitons in multiple reservoirs $n(\mathbf{r})$ can be written as:
\begin{align}
    \frac{dN_\nu(\mathbf{r})}{dt}&=\sigma n(\mathbf{r})N_\nu(\mathbf{r})-\frac{N_\nu(\mathbf{r})}{\tau} \nonumber \\
    \frac{dn(\mathbf{r})}{dt}&=P(\mathbf{r})-\gamma n(\mathbf{r})-\sigma n(\mathbf{r})N_\nu(\mathbf{r}).
\end{align}
In the following, we consider a continuous wave pumping regime where the exciton reservoir density can be assumed independent of time, $\frac{dn(\mathbf{r})}{dt}\approx0$. This will lead to the following important relation between the condensate and reservoir densities:
\begin{equation}
    n(\mathbf{r})=\frac{P(\mathbf{r})}{\gamma+\sigma N_\nu(r)}.
\end{equation}
Substituting this equation into $\frac{dN_\nu(\mathbf{r})}{dt}$, we obtain:
\begin{equation}
    \frac{dN_\nu(\mathbf{r})}{dt}=\frac{\sigma P(\mathbf{r})N_\nu(\mathbf{r})}{\gamma+\sigma N_\nu(r)}-\frac{N_\nu(\mathbf{r})}{\tau}.
\end{equation}
Integrating over the space, 
\begin{equation}
    \frac{dN_\nu}{dt}=-\frac{N_\nu}{\tau}+\int \frac{\sigma P(\mathbf{r})N_\nu(\mathbf{r})}{\gamma+\sigma N_\nu(r)} d\mathbf{r}.
\end{equation}
Now we substitute the Hankel ansatz for the single-condensate wavefunction $\Psi(\mathbf{r})=H^{(1)}_0(k_c|\mathbf{r-\mathbf{r_i}}|)$, and $N_\nu(\mathbf{r})=|\Psi(\mathbf{r})|^2$. In the numerical modeling, we shall also assume a Gaussian pump (that is the most frequently used experimentally) $P(\mathbf{r})=\exp(-|\mathbf{r}-\mathbf{r}_i|^2/2w^2)$, where $w$ is the width of the pump spot. Then the kernel of the integral is a Hankel function modulated by a Gaussian profile (see Appendix B for details).

Notice that, if $\gamma=0$, which means the exciton reservoir has an infinite lifetime, this equation is reduced to one describing the case of resonant pumping:
\begin{equation}
    \frac{dN_\nu}{dt}=-\frac{N_\nu}{\tau}+P.
\end{equation}
Due to the radiative decay of exciton-polaritons given by $-N_\nu/\tau$, the mode with the smallest eigenvalue is characterized with the lowest threshold pump power of the Bose-Einstein condensation. 



In the experiments of Ref \cite{berloff2017realizing}, the size of the pump spot is 1-2 um, compared to the large lattice constant $a>10$um, the size of the pump spot can be neglected and treated as a Dirac $\delta$-function. In this way, $P(\mathbf{r-r_i})\approx P_0\delta(\mathbf{r-r_i})$. Then the rate equation for the population of a multi-condensate eigen mode can be simplified as:
\begin{align*}
    \frac{dN_\nu}{dt}&=-\frac{N_\nu}{\tau}+\int \frac{\sigma P_0\delta(\mathbf{r}) N_\nu(\mathbf{r})}{\gamma+\sigma N_\nu(\mathbf{r})} d\mathbf{r}\\
    &=-\frac{N_\nu}{\tau}+\Sigma_i\sigma\frac{P_0N_\nu(\mathbf{r_i})}{\gamma+\sigma N_\nu(\mathbf{r_i})}
\end{align*}
where $\mathbf{r_i}$ corresponds to pump spot at the $i$-th lattice site. This function, $\frac{\sigma P_0}{\gamma/\Sigma_iN_\nu(\mathbf{r_i})+\sigma }$ increases monotonically with $\Sigma_iN_\nu(\mathbf{r_i})$. This implies that the modes with the largest eigenvalue has the steepest slope of the growth rate. Together with the homogeneous decay $-N_\nu/\tau$, the mode selection mechanism can be interpreted as follows: at the low pump power, the mode with the minimum overlap between all individual polariton condensates will dominate; once the pump power increases,  the population of modes with higher overlaps start growing faster. Eventually, at the high pump power limit, the mode with the maximum overlap between wavefunctions of individual polariton condensates will exhibit the highest population growth rate. Most likely, this mode will be the only one to survive at a long time-scale.


\section{Mapping to the XY model}

The overlap matrix $D$ introduced in Eq.~\ref{Eq:OME} has a form
\begin{equation}
\mathcal{D}=\sum_i\epsilon_0c_i^\dagger c_i+\sum_{\langle i,j\rangle}\left(\beta_{ij}c_i^\dagger c_j+\beta_{ij}^*c_j^\dagger c_i\right),
\end{equation}
where $\epsilon_0=1$, $\beta_{ij}=D_{ij}$, and $c_i$ is the annihilation operator corresponding to the state $\ket{\Psi_i}$. Mathematically, the overlap matrix $D$ is isomorphic to a tight-binding Hamiltonian with an on-site energy $\epsilon_0$. This allows us to draw an analogy between overlap integrals $\beta_{ij}$ and amplitudes of hopping between lattice sites:
\begin{equation}
\mathcal{H}=\sum_i\epsilon_0c_i^\dagger c_i+\sum_{\langle i,j\rangle}\left(\beta_{ij}c_i^\dagger c_j+\beta_{ij}^*c_j^\dagger c_i\right),
\end{equation}

To map an array of phase-locked exciton-polariton condensates to a spin-$1/2$ XY model, one can use the Holstein–Primakoff transformation\cite{holstein1940field,nieto1997holstein} in the hard-core limit:
\begin{equation}
c_j^\dagger\leftrightarrow S_j^+,\quad c_j\leftrightarrow S_j^-,\quad c_j^\dagger c_j\leftrightarrow S_j^z+\frac{1}{2},
\end{equation}
where $S_j^\pm=S_j^x\pm iS_j^y$ are the spin raising and lowering operators. The bosonic hopping term can be rewritten as:
\begin{equation}
c_i^\dagger c_j+c_j^\dagger c_i=S_i^+S_j^-+S_i^-S_j^+.
\end{equation}
Using the identity:
\begin{equation}
S_i^+S_j^-+S_i^-S_j^+=2(S_i^xS_j^x+S_i^yS_j^y)
\end{equation}
we obtain the spin representation of the hopping Hamiltonian:
\begin{equation}
\mathcal{H}_\mathrm{hop}=2\sum_{\langle i,j\rangle}\left(D_{ij}S_i^xS_j^x+D_{ji}S_i^yS_j^y\right).
\label{Hhop}
\end{equation}
Bearing in mind that the standard classical XY Hamiltonian is defined as:
\begin{equation}
\mathcal{H}_{\mathrm{XY}}=\sum_{\langle i,j\rangle}\left(J_{ij}S_i^xS_j^x+J_{ji}S_i^yS_j^y\right),
\end{equation}
where $J$ is the exchange coupling. It can be seen that the Hamiltonian $\mathcal{H}_\mathrm{hop}$ is of exactly the same form as the classical XY model Hamiltonian for in-plane spin components.
We identify the effective exchange coupling strength as
$J_{ij}=2D_{ij}$, where $D_{ij}$ are the elements of the matrix of overlap integrals of single-condensate wavefunctions introduced above (see Eq.~\ref{Eq:OME}. If $D_{ij}$ are real numbers, two important particular cases can be observed.
\begin{itemize}
\item If $D_{ij}<0$, then $J_{ij}<0$, which corresponds to the ferromagnetic coupling in the XY model.
\item If $D_{ij}>0$, then $J_{ij}>0$, which corresponds to the antiferromagnetic coupling.
\end{itemize}
In a general case, however, the matrix elements $D_{ij}$ are complex numbers. Consequently, the Hamiltonian $\mathcal{H}_{\mathrm{XY}}$ can describe frustrated phases, a spin glass phase etc.
The remarkable mapping between eigenvectors of the matrix $D$ and the eigenstates of the Hamiltonian $\mathcal{H}_{\mathrm{XY}}$ is at the core of the operational principle of the polariton XY simulators. Note that our model fully accounts for the coherent and dissipative coupling mechanism of exciton-polariton condensates\cite{nalitov2019optically}. The former one is accounted for in the off-diagonal elements of the overlap matrix $D_{ij}$, while the latter one is introduced in the rate equations of multi-condensate modes. It is the dissipative coupling that causes the system to choose one of the eigenmodes of the matrix $D$.

\section{Numerical results}

Here we consider three important specific examples of phase locking in arrays of exciton-polariton condensates, namely (I) a linear chain, (II) a triangular lattice, (III) a random graph. We shall use the following set of parameters: $a=10\ \mu \mathrm{m}$ being the spacing between neighboring pump spots in the cases (I) and (II) and the average spacing between pump spots in the case (III), $\tau=1$ ps, and $1/\gamma=0.1$ ps, $\sigma=0.01\ \mu \mathrm{m}^{2}\mathrm{ps}^{-1}$. Below we show the growth rate $\frac{dN_\nu}{dt}$ for the modes of minimum eigenvalue and maximum eigenvalue, i.e. $N_1$ and $N_N$. The typical result is shown in Fig.~\ref{fig_pump} which refers to a $10\times10$ triangular lattice (similar population growth rates are observed for the $33\times33$ lattice, as discussed later). It can be seen that the mode with the smallest eigenvalue (blue curves) is characterized with the smallest threshold pump power and the smallest slope of the growth rate plotted versus pump power. This implies that when the pump power is small, the mode characterized by the lowest overlap of single-condensate wavefunctions will be the first to form an extended macroscopically occupied polariton condensate. With the increase of the pump power, the modes with higher eigenvalues grow faster, and eventually the mode corresponding to the highest eigenvalue of the matrix $D$ will win in the competition. Saying so, we must realize that some of the eigenmodes may be degenerate, dependent on the specific shape of the polariton graph. The competition between degenerate or nearly degenerate modes may cause bi-stability, multi-stability and a stochastic behavior. These effects remain beyond the scope of the present study. 

\subsection{A linear chain}

For a periodic one-dimensional chain of polariton condensates, the effective Hamiltonian describing coherent coupling (hopping) between neighbor sites can be written as:
\begin{equation}
\mathcal{H}=\sum_i\epsilon_0c_i^\dagger c_i+\sum_{i}\left(\beta_{i,i+1}c_i^\dagger c_{i+1}+\beta_{i+1,i}c_{i+1}^\dagger c_i\right),
\end{equation}
where $\beta_{i,i+1}=D_{i,i+1}$ is the hopping amplitude between sites $i$ and $i+1$, and $\epsilon_0=1$ is the on-site energy. The resulting eigen-energy for the mode characterized by a Bloch wavevector $\mathbf{k}$ is:
\begin{equation}
E(\mathbf{k})=1+(\beta+\beta^*)\cos{ka}
\end{equation}
where $\beta$ is the nearest-neighbor overlap integral. We remind that $a$ is the lattice constant equal to the distance between neighboring pump spots. The extreme values of the energy dispersion $E(\mathbf{k})$ determine the ground state of the system:
\begin{itemize}
    \item The eigenvalue $E_+$ occurs at the Brillouin zone center $k=0$:
    \begin{equation}E_+=1+2\mathrm{Re}(\beta)\end{equation}
    \item The eigenvalue $E_-$ occurs at the edges of the Brillouin zone ($k=\pi/a$):
    \begin{equation}E_-=1-2\mathrm{Re}(\beta)\end{equation}
\end{itemize}
The former state corresponds to a ferromagnetic (FM) phase where all condensates have the same phase. The latter state corresponds to an anti-ferromagnetic phase (AFM) where the orientations of effective spins of the neighboring condensates differ by 180°.

\begin{figure*}
    \centering
    \includegraphics[width=1\linewidth]{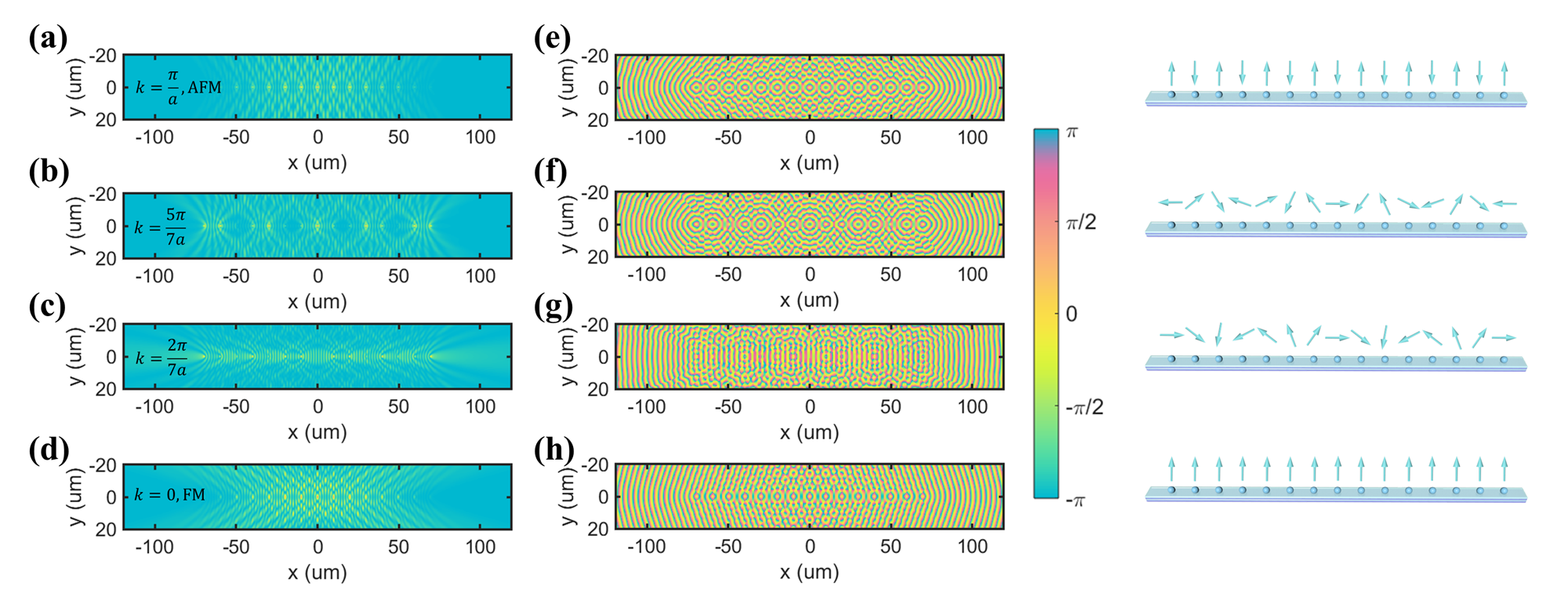}
    \caption{(a-d) Spatial distributions of polariton densities characterizing the multi-condensate eigenmodes in a one dimensional chain of polariton condensates for $N_1$ mode with $k=\pi/a$, $N_5$ mode with $k=5\pi/7a$, for $N_{10}$ mode with $k=2\pi/7a$, and for $N_{15}$ mode with $k=0$. (e-h) Corresponding phase patterns and schemes which show the orientations of effective spins obtained by mapping the system to an XY-Hamiltonian. The calculation is performed assuming $k_c=2\ \mu \mathrm{m}^{-1}$.}
    \label{fig_1Dchain}
\end{figure*}

The numerical result for the one dimensional polariton chain of 15 condensates obtained assuming $k_c=2\ \mu \mathrm{m}^{-1}$ is shown in Fig.~\ref{fig_1Dchain}. The mode with the smallest eigenvalue $N_1$ exhibits an anti-ferromagnetic phase, corresponding to the case where the Bloch wavevector $k=\pi/a$. In contrast, the mode with the largest eigenvalue $N_{15}$ exhibits the ferromagnetic alignment. The phase difference between nearest neighbor condensates of the intermediate modes is given by $\Delta\theta=(\nu-1)\pi/14-\pi$, where $\nu$ is the index of the corresponding eigenmode.

\subsection{A triangular lattice}

\begin{figure}
    \centering
    \includegraphics[width=1\linewidth]{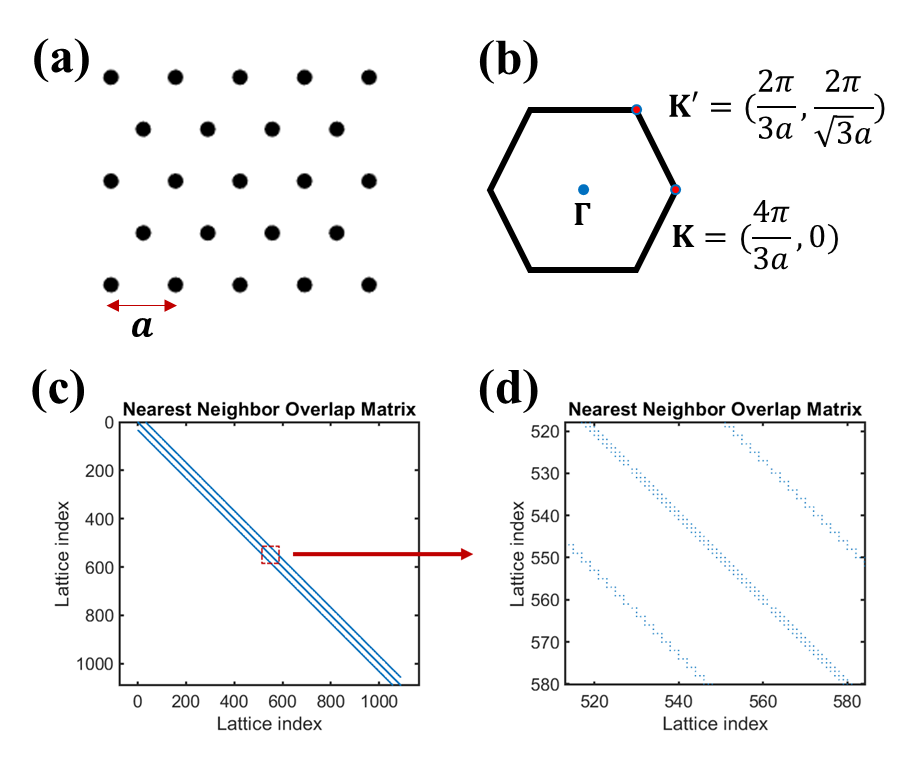}
    \caption{(a) A schematic of pump power distribution in real space that is used to create a zigzag triangular polariton lattice, characterized by the lattice constant $a$. (b) The first Brillouin zone of the corresponding lattice of polariton condensates. At $\Gamma$ point the eigenvalue of the overlap matrix $D$ is $E_+(\mathbf{k})$, while at two-fold degenerate $K$ and $K^\prime$ points the corresponding eigenvalue is $E_-(\mathbf{k})$. (c) The structure of the sparse overlap matrix for an open-boundary $33\times 33$ zigzag triangular lattice. (d) A magnified structure of the overlap matrix.}
    \label{fig_lattice}
\end{figure}

\begin{figure}
    \centering
    \includegraphics[width=1\linewidth]{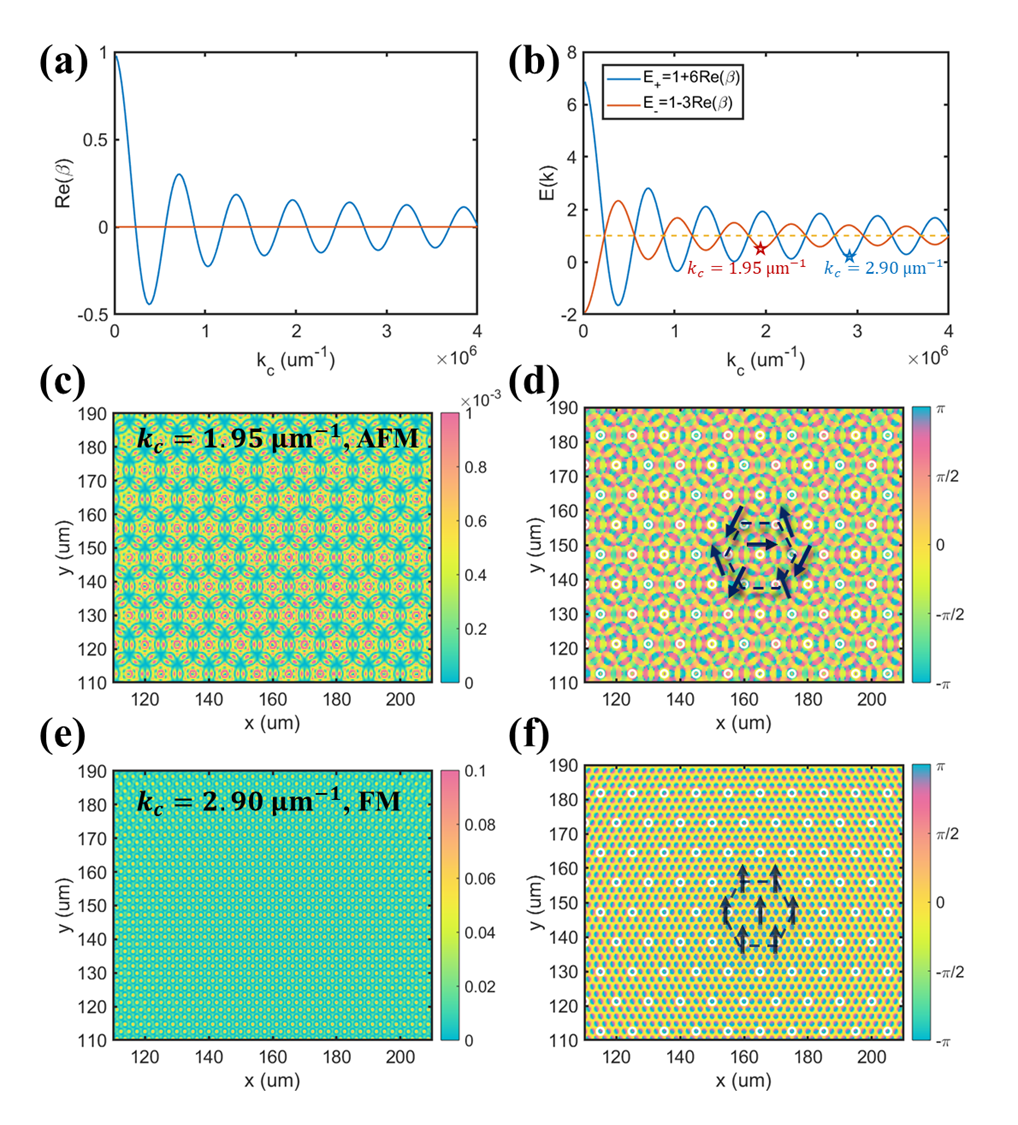}
    \caption{(a) The dependence of the overlap integral $\mathrm{Re}(\beta)$ on $k_c$. The sign of $\mathrm{Re}(\beta)$ determines the ground state of the polariton lattice. (b) The dependence of the extreme values of the eigenenergies $E_\pm$ on the wavevector $k_c$. The lowest eigenvalue of the system (below the yellow dashed line) is either $E_+$ ($\Gamma$-point) or $E_-$ ($K$-point), depending on the value of $k_c$, which corresponds to the ferromagnetic phase or frustrated anti-ferromagnetic phase, respectively. (c) The spatially dependent polariton density $|\varphi|^2$ of the multi-condensate eigenmode at $k_c=1.95\ \mathrm{\mu m}^{-1}$ and (d) the corresponding phase pattern. The phases of the condensates at neighboring sites differs by $2\pi/3$, indicating a frustrated anti-ferromagnetic phase. The white circles in the figure below represent the lattice sites where the optical pump beams are focused. Directions of effective spins for each individual condensate are shown on a selected hexagonal unit cell. (e) The polariton density $|\varphi|^2$ of the multi-condensate eigenmode at $k_c=2.90\ \mathrm{\mu m}^{-1}$ and (f) the corresponding phase pattern. Here phases of the condensates at different lattice sites are the same, which indicates a ferromagnetic phase. Directions of the effective spins corresponding to each individual condensate are shown by arrows on a selected hexagonal unit cell.}
    \label{fig_phase}
\end{figure}

For a periodic triangular lattice (see Fig.\ref{fig_lattice}a), the problem of diagonalization of the overlap matrix maps directly onto a tight-binding model. The effective Hamiltonian describing coherent coupling (hopping) between sites is:
\begin{equation}
\mathcal{H}=\sum_i\epsilon_0c_i^\dagger c_i+\sum_{\langle i,j\rangle}\left(\beta_{ij}c_i^\dagger c_j+\beta_{ij}^*c_j^\dagger c_i\right),
\end{equation}
where $\beta_{ij}=D_{ij}$ is the hopping amplitude between sites $i$ and $j$, and $\epsilon_0=1$ is the on-site energy. For a translationally invariant system, we apply the Bloch's theorem. The wavefunction on a lattice site at $\mathbf{r}_i$ takes the form:
\begin{equation}
\psi_\mathbf{k}(\mathbf{r}_i)=\frac{1}{\sqrt{N}}e^{i\mathbf{k}\cdot\mathbf{r}_i}
\end{equation}
The resulting eigen-energy for the mode characterized by the Bloch wavevector $\mathbf{k}$ is:
\begin{equation}
E(\mathbf{k})=1+(\beta+\beta^*)\gamma(\mathbf{k})
\end{equation}
where $\beta$ is the nearest-neighbor overlap integral
and $\gamma(\mathbf{k})$ is the geometric structure factor for the triangular lattice:
\begin{equation}
\gamma(\mathbf{k})=\cos(k_xa)+2\cos\left(\frac{k_xa}{2}\right)\cos\left(\frac{\sqrt{3}k_ya}{2}\right).
\end{equation}
$a$ is the lattice constant. The corresponding Bloch wavefunction for the entire system is a coherent superposition of all single-condensate wavefunctions:
\begin{equation}
\varphi_\mathbf{k}(\mathbf{r})=\frac{1}{\sqrt{N}}\sum_{i=1}^Ne^{i\mathbf{k}\cdot\mathbf{r}_i}H_0^{(1)}(k_c|\mathbf{r}-\mathbf{r}_i|).
\end{equation}

The extreme values of the energy dispersion $E(\mathbf{k})$ determine the ground state of the system:
\begin{itemize}
    \item The eigenvalue $E_+$ corresponds to the Brillouin zone center (the $\Gamma$-point):
    \begin{equation}E_+=1+6\mathrm{Re}(\beta)\end{equation}
    This state corresponds to a ferromagnetic (FM) phase where all individual condensates have the same phase.
    \item The eigenvalue $E_-$ corresponds to the corners of the Brillouin zone ($K$-points):
    \begin{equation}E_-=1-3\mathrm{Re}(\beta)\end{equation}
\end{itemize}
This state corresponds to a frustrated anti-ferromagnetic phase, where the phase of the eigenmode exhibits a $2\pi/3$ shift between neighboring sites, and the effective spins characterizing individual condensates within the XY-model are rotated by 120° with respect to each other. 

The sign and the magnitude of the real part of the overlap integral $\mathrm{Re}(\beta)$ oscillate as functions of $k_ca$, as shown in Fig.\ref{fig_phase}a. Consequently, the global ground state of the system is determined by the competition between $E_+$ and $E_-$. By tuning the polariton energy (and thus $k_ca$) via, e.g., exciton-photon detuning that can be controlled by external electric or magnetic field, the system can be switched between ferromagnetic and antiferromagnetic order, realizing a programmable XY simulator.

To consider an example of a large-size polariton XY simulator, we have diagonalized the $1089 \times 1089$ overlap matrix $D$ for a $33\times 33$ triangular lattice. The spectrum of eigenvalues of this matrix reproduces the tight-binding band structure. The numerical diagonalization of the overlap matrix and the subsequent analysis of the eigenstates provide a clear picture of the emergent magnetic order in a an array of two-dimensional classical spins arranged in a triangular lattice. Our analysis allows linking the eigenvalues of the effective Hamiltonian to the observed phase patterns and it predicts the dynamics of their build-up. Fig.\ref{fig_phase}b shows that the nature of the ground state in our polariton lattice can be dynamically selected by tuning in-plane wavevector $k_c$, e.g. by application of external fields.

For the values of $k_c$ corresponding to the negative real part of the overlap integral (e.g., $k_c=1.95\ \mathrm{\mu m}^{-1}$), the minimum eigenvalue corresponds to the $\Gamma$-point. As shown in Fig.\ref{fig_phase}c, this results in a phase-locked state where all condensates oscillate with the same phase. This uniform phase configuration is the direct analog of the ferromagnetic spin order in the XY model, where all spins are aligned. 

For the values of $k_c$ leading to the positive values of $\beta$ (e.g., $k_c=2.90\ \mathrm{\mu m}^{-1}$, the ground state shifts to the $K$-points of the Brillouin zone. The corresponding eigenstate, visualized in Fig.\ref{fig_phase}d, exhibits the hallmark 120° spin order. In this configuration, the phases of three neighboring condensates forming an elementary triangle differ from each other by $2\pi/3$. This is the well-known, geometrically frustrated ground state of the antiferromagnetic XY-model on a triangular lattice. 
The tunability of the system between FM and 120° AFM phases by varying $k_c$ underscores the flexibility of the polariton lattice for XY-type simulations, allowing for the exploration of different magnetic regimes in a single, controllable platform.

\subsection{A random graph}
\begin{figure}
    \centering
    \includegraphics[width=1\linewidth]{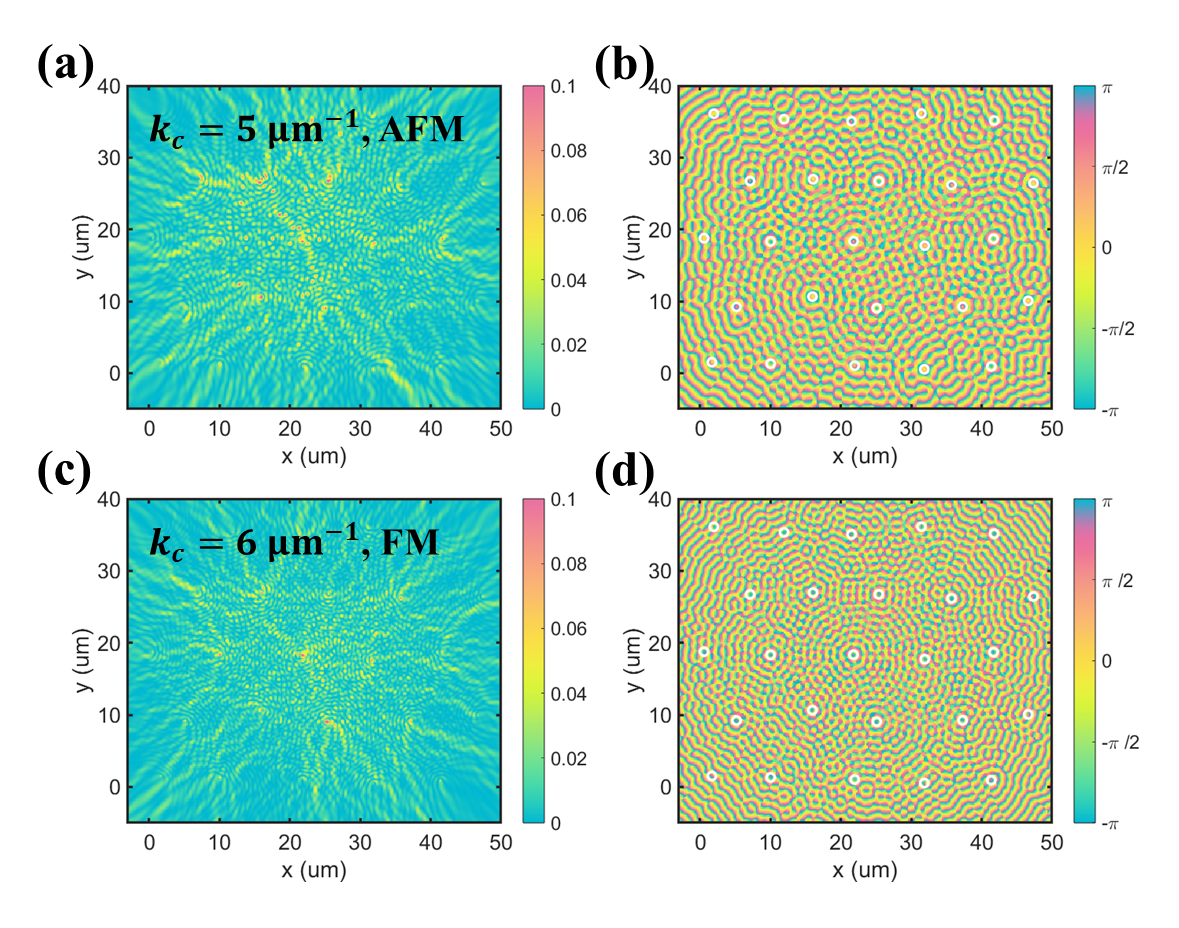}
    \caption{Predictions of the spin structure for a randomized $5\times5$ zigzag triangular polariton lattice provided by a polariton XY-simulator. (a), (b) The intensity and phase maps of the frustrated AFM eigenstate for the disordered polariton lattice at $k_c=5\ \mu \mathrm{m}^{-1}$, respectively. (c), (d) The intensity and phase maps of the frustrated AFM eigenstate for the disordered polariton lattice at $k_c=6\ \mu \mathrm{m}^{-1}$, respectively.}
    \label{fig_random}
\end{figure}

In the preceding sections, our analysis of the solutions of the XY Hamiltonian problem relied on the perfect translational symmetry of an ideal lattice of polariton condensates, which ensured that the overlap matrix element $\braket{\Psi_j}{\Psi_i}$ was the same for all pairs of neighbors. In contrast, for a random graph, the overlap matrix becomes explicitly position-dependent, lacking the strict periodicity of the ideal case. Consequently, the concept of Bloch waves and well-defined energy bands is no longer applicable, generally. 
Clearly, in the case of strong randomness, the modes characterized by the lowest threshold pumping or the highest growth rate will not necessarily be characterized by FM or AFM spin alignment. 
Still, if deviations from the translations symmetry are sufficiently weak, such as in the spin-glass model, where a short-range order is maintained, the mode selection mechanism based on the analytical model developed here is still valid. We prove this statement by introducing small random perturbations to the site positions in the $5\times5$ triangular polariton lattice with a lattice constant $a=10\ \mu \mathrm{m}$. The result of this simulation is shown in Fig.~\ref{fig_random}. By tuning the wavevector $k_c$, which controls the phase of the coherent coupling between sites, we can still clearly observe the transition from FM phase (all condensates phase-locked at zero relative phase) to a frustrated AFM phase (phase difference between nearest neighbors remains approximately equal to $2\pi/3$).

The persistence of these phases underscores the robustness of magnetic order in XY-spin arrays. As long as the disorder is weak enough so that the local coupling between nearest neighbors is not fundamentally altered, the system can still settle into ordered phases of the effective XY model. This resilience is highly promising for experimental realizations, where the perfect periodicity is often difficult to achieve. 

\section{Pump power dependence}

In the previous sections, we demonstrated that the pump power $P$ eventually determines which one of the eigenmodes of the effective XY Hamiltonian will be selected by the system. In order to find the critical pump power needed to select a specific mode $\varphi_\nu(\mathbf{r})$ and make sure that it will exhibit the fastest population growth rate, one may need to account for the dependence of the single condensate wavefunction parameter $k_c$ on the pump power. In the previous consideration, $k_c$ was taken as an independent parameter for simplicity. However, in a general case it is power-dependent. For exciton-polariton condensates formed under the pump spots, the value of $k_c$ is governed by the balance of the potential energy of polaritons interacting with the exciton reservoir that scales as $n_0/S$, where $n_0$ is the total population of the reservoirs and $S$ is the area of the system, and the kinetic energy of exciton-polaritons ballistically propagating between the condensates that scales as $k_c^2$. The square root dependence of $k_c$ on $n_0/S$ manifests itself in the variation of the overlap integrals of individual exciton-polariton condensates that affects the effective Hamiltonian matrix, its eigenvalues and eigenvectors.

\section{Conclusion}
We have developed an analytical model linking the matrix of overlap integrals of individual polariton condensates $D$ to an effective XY Hamiltonian. We have demonstrated that in an array of $N$ coupled polariton condensates formed by non-resonant optical pumping, $N$ stable phase-locked solutions correspond to the eigenmodes of the matrix $D$. It depends on the pump power which one of these N states will be chosen by the system. We have shown that the the mode with the lowest eigenvalue of the matrix $D$ is characterized by the lowest threshold to Bose-Einstein condensation. Gradually increasing the pump power above threshold, one can make any of the other modes preferential from the point of view of the rate of the condensate formation. In the high power limit, the mode corresponding to the largest eigenvalue of the matrix $D$ would win.

This surprisingly simple scenario of phase-locking in arrays of exciton-polariton condensates can be understood following a simple intuition based on two assumptions: (1) the mode characterized by the lowest rate of radiative losses requires the lowest pump power to achieve the bosonic condensation threshold, (2) in contrast, the mode with a highest threshold power demonstrates the highest growth rate at threshold. The latter follows from the fact that the reservoir occupation number growth proportionally to the pump power and the condensate growth rate depends on the product of its occupation number and the reservoir occupation number.

Obviously, this simplistic picture can only be resorted to in a narrow range of pump powers, in the vicinity of bosonic condensation thresholds in the system. Fortunately, the behavior of the system in this narrow pump range is the most relevant to the operation of polariton XY simulators. Indeed, once a macroscopically occupied coherent multicondensate state is formed, it is getting stabilized by the stimulated scattering, which makes phase locking in arrays of polariton condensates surprisingly stable \cite{berloff2017realizing}. Still, special regimes such as the limit cycles or stochastic behavior caused by multi-stability are possible. The applicability of a polariton XY simulators in these (rather exotic) regimes would not be possible.

Here we have considered three specific geometries of arrays of coupled polariton condensates, where XY simulations work with a high accuracy. Namely, we studied a linear chain, a triangular lattice and a random graph. In all these cases, tuning the pump power one can visit the antiferromagnetic and ferromagnetic phase configurations as well as irregular spin glass modes.


In is very important to note that a desired eigen-state of the effective Hamiltonian can be reached by the system on an ultrafast timescale (on the order of the single condensate formation time that is a few tens of picoseconds)\cite{topfer2020time,harrison2020synchronization}. This highlights a significant advantage of polariton simulators over iterative computational methods. In particular, the potential of polariton lattices as high-speed, analog simulators for complex classical magnetism is very high. This work establishes a physical bridge between polariton condensation and classical frustrated spin models, paving the way for programmable XY spin simulators operating at the picosecond timescale. It rectifies shortcomings of the previous models in what concerns the competition between phase-locked configurations corresponding to different eigen-states of the effective XY Hamiltonian.

\begin{acknowledgements}
AVK and JC thanks Pavlos Lagoudakis, Yuriy Rubo and Anton Nalitov for stimulating discussions. AVK acknowledges support from Saint Petersburg State University (Research Grant No. 125022803069-4) and from the Innovation Program for Quantum Science and Technology (No. 2021ZD0302704). 
\end{acknowledgements}

\appendix

\section{Derivation of the relation $N_\nu=\lambda_\nu$}

This relation is important for the interpretation of the dependence of the threshold power of the multi-condensate eigenmodes on the corresponding eigenvalues of the overlap matrix $D$.
We start from
\begin{align}
N_\nu&=\int|\varphi(\boldsymbol{r})|^{2}d\boldsymbol{r} \nonumber \\ \nonumber
&=\int\Sigma_{j}\frac{1}{\sqrt{N}}e^{-i\mathbf{k}\cdot\mathbf{r}_{j}}\Psi_{j}^{*}(\boldsymbol{r})\Sigma_{i}\frac{1}{\sqrt{N}}e^{i\mathbf{k}\cdot\mathbf{r}_{i}}\Psi_{i}(\boldsymbol{r})d\boldsymbol{r} \\ \nonumber
&=\frac{1}{N}\Sigma_{ij}e^{-i\mathbf{k}\cdot(\mathbf{r}_j-\mathbf{r}_i)}\int\Psi_j^*(\boldsymbol{r})\Psi_i(\boldsymbol{r})d\boldsymbol{r}.\\ \nonumber
\end{align}
For polariton lattice characterized by a translational symmetry, one can apply the Bloch theorem, that allows representing the eigenstate of the system as: $c_j=e^{i\mathbf{k}\cdot\mathbf{r}_j}$. Then, the secular equation becomes:
\begin{equation}
\Sigma_iD_{ij}e^{i\mathbf{k}\cdot\mathbf{r}_i}=\lambda e^{i\mathbf{k}\cdot\mathbf{r}_j}.
\end{equation}
Using the equality of the nearest neighbor coupling one can obtain
\begin{equation}
   \frac{1}{N}\sum_{ij}e^{-ik\cdot(r_j-r_i)}D_{ij}=\sum_Re^{-ik\cdot R}D_{ij} 
\end{equation}
where $R=|r_i-r_j|$. Notice that $\sum_Re^{-ik\cdot R}D_{ij}=\lambda$. Therefore, $N_\nu=\lambda_\nu$ for the specific mode $\nu$.

We shall mention that, in our formalism, the eigenvalue $\lambda_\nu$ of the overlap matrix depends on the normalization factor (or the effective particle number) associated with the chosen basis set. As an example, if the single-condensate basis functions are chosen as Hankel functions, they are not delocalized and cannot be strictly orthogonal. This may result in a non–positive-definite overlap matrix. Consequently, some eigenvalues of the overlap matrix may become negative (extremely, when $k_c$ is small), reflecting the non-orthogonality and oscillatory nature of the Hankel basis. Crucially, a negative $\lambda_\nu$ does not correspond to a negative particle number in a physical sense. Instead, it signifies that the corresponding eigenmode of the overlap matrix is not a valid, normalized quantum state within the chosen non-orthogonal basis. The relation $N_\nu=\lambda_\nu$ should be understood as an algebraic identity stemming from the plane-wave ansatz and the structure of the overlap matrix. Physically, the total particle number for any mode must be positive. Hence the use of the Hankel ansatz for the single-condensate basis is only valid if the overlap matrix is positively defined (when $k_ca$ is large enough, to ensure the nearest neighbor coupling dominates). The appearance of negative eigenvalues is thus an artifact of the specific basis set and its inherent non-orthogonality. In our numerical simulations for the realistic arrays of polariton condensates, the minimum eigenvalue is found to be positive for the dominant, low-energy modes considered, ensuring the correctness of the choice of the basis as well as the physical consistency of our results.

\section{Effect of finite size of the pump spot}
In the main text, it is assumed that the size of the pump spot is small and therefore it can be treated as a $\delta$-function. This approximation is valid of the individual condensates are sufficiently far from each other which is typically the case in the published experiments \cite{berloff2017realizing}.
However, in a general case, the ratio $w/a$ between the width of the pump spot $w$ and the lattice constant $a$ may be non-negligible. A single-condensate wavefunction can be conveniently represented as a convolution of the reservoir intensity profile and the Hankel function describing the emission from a point source in this case.
$\varphi(\mathbf{r})$ is the integral of all possible paths with the propagator $G(\mathbf{r-r^\prime})$:
\begin{align}
\varphi(\mathbf{r})&=\int G(\mathbf{r}-\mathbf{r}^{\prime})i\frac{2mR}{\gamma}P(\mathbf{r})\varphi(\mathbf{r}^{\prime})d\mathbf{r}^{\prime}\\
&=\int-\frac{1}{4}H_0^{(1)}(k_0|\mathbf{r}-\mathbf{r}^{\prime}|)\frac{2mR}{\gamma}P(\mathbf{r})\varphi(\mathbf{r}^{\prime})d\mathbf{r}^{\prime}\\
&=A\int H_0^{(1)}(k_0|\mathbf{r}-\mathbf{r}^{\prime}|)P(\mathbf{r})\varphi(\mathbf{r}^{\prime})d\mathbf{r}^{\prime}
\end{align}
where $A=-\frac{mR}{2\gamma}$ is a constant,  $k_0^2=2m(\mu+i\gamma_c)$, $\mu$ is the chemical potential, and $P(\mathbf{r})$ is the profile of the pump. This integral equation can be numerically calculated using the convolution theorem by iteration method.

\begin{figure}
    \centering
    \includegraphics[width=1\linewidth]{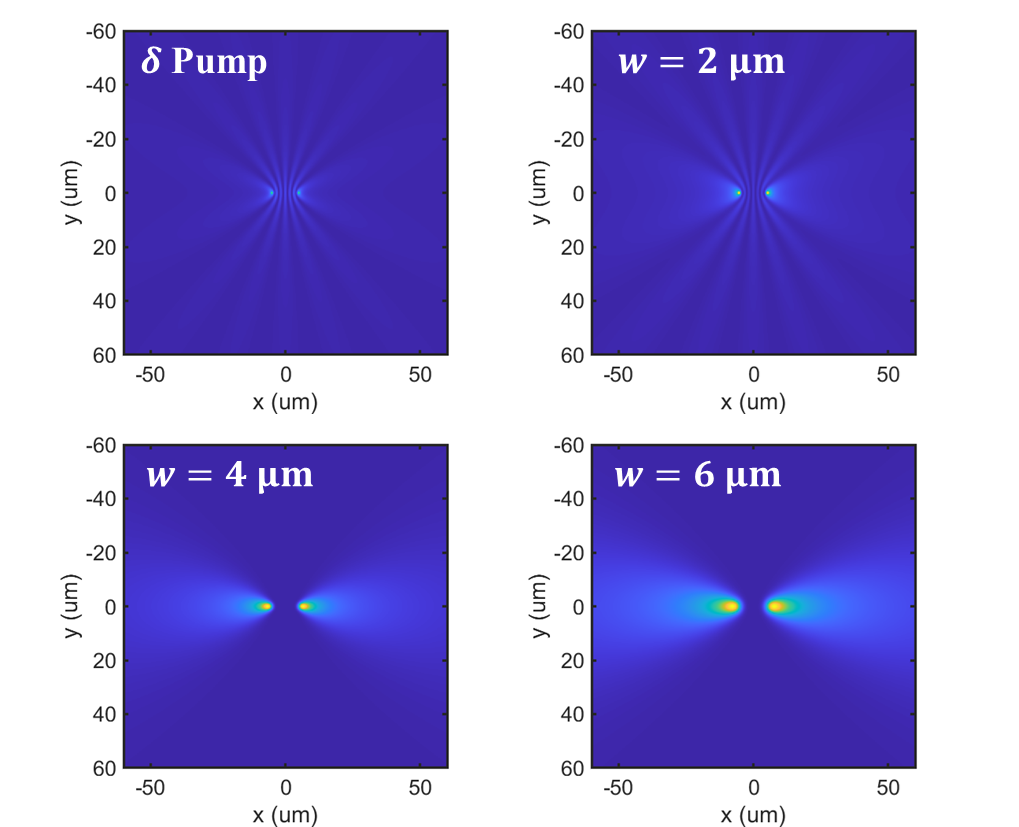}
    \caption{Interference patterns of two polariton condensates created by pump beams of different lateral sizes $w$.}
    \label{fig_si_pz}
\end{figure}

The impact of the pump spot size on the coherence of polariton condensates is demonstrated in Fig.~\ref{fig_si_pz}, which shows the interference patterns for two condensates spaced $a=10\ \mu\text{m}$ apart and emitting polaritons characterized with an in-plane wavevector $k_0=2\ \mu\text{m}^{-1}$. These patterns are generated under Gaussian pumps $P(\mathbf{r})=P_0\exp(-r^2/2w^2)$ of different widths $w$. In the limiting case of a $\delta$-function pump, each single-condensate wavefunction is described by a Hankel function, leading to a well-defined interference pattern. This interference remains observable for a small pump spot ($w=2\ \mu\text{m}$), confirming that the Hankel function convolution with the reservoir density distribution still preserves the oscillatory tails of the wavefunctions beyond the pump spots. In contrast, in the case of large pump spots $w>4\ \mu \mathrm{m}$ the interference is suppressed. Spatially extended pumps introduce a significant spread in phase of polaritons emitted in radial directions, which impacts on the visibility of interference fringes and, therefore, affects the overlap integrals that constitute off-diagonal elements of the matrix $D$. Still, the phase-locked multi-condensate eigenstates can be found by diagonalization of the overlap matrix $D$, and their dynamics can be analyzed as described above. None of the qualitative results of the model presented here would be changed.

\begin{figure}
    \centering
    \includegraphics[width=1\linewidth]{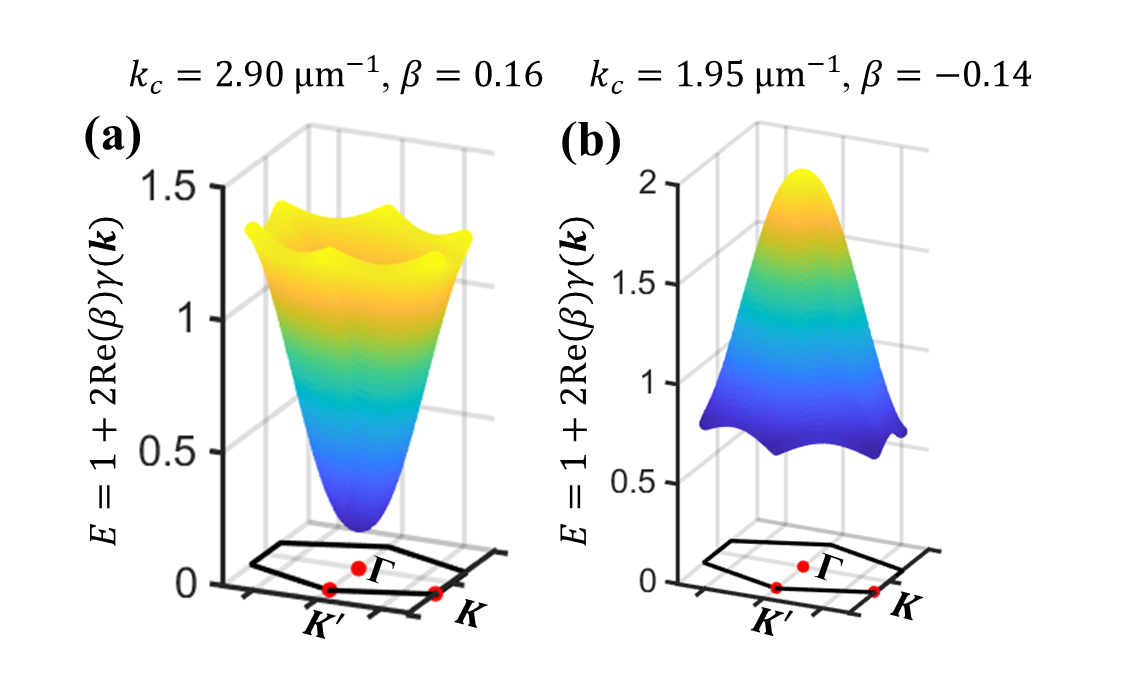}
    \caption{Eigenvalue for the effective Hamiltonian of the triangular polariton lattice at (a) $k_c=2.90\ \mu \mathrm{m}^{-1}$ and (b) $k_c=1.95\ \mu \mathrm{m}^{-1}$.}
    \label{fig_si_ev}
\end{figure}

\section{The eigenvalue spectrum in a triangular lattice}
In the context of a triangular polariton lattice, the ground state of the system is determined by the sign of the overlap matrix element $\beta$, as follows from the eigenvalue dispersion $E(\mathbf{k})=1+2\mathrm{Re}(\beta)\gamma(\mathbf{k})$ for the two scenarios depicted in Fig.~\ref{fig_phase}.

In Fig.~\ref{fig_si_ev}(a), a positive $\beta$ results in the minimum eigenvalue occurring at the $\Gamma$ point. This signifies that the Bloch wavefunction exhibits a uniform phase across all lattice sites, leading to the formation of a ferromagnetic order.

Conversely, in Fig.~\ref{fig_si_ev}(b), a negative $\beta$ indicates that the system favors an anti-phase alignment between neighboring sites. Consequently, the ground state of the XY simulator is an anti-ferromagnetic order, which, due to geometric frustration on the triangular lattice, is realized at the $K$ and $K'$ points in momentum space. The two-fold degeneracy at these points corresponds to the two possible chiral spin configurations with phase shifts of $-2\pi/3$ and $2\pi/3$.

Furthermore, the mode selection dynamics under non-resonant pumping can be understood as follows. For positive $\beta$, the $\Gamma$-point mode, possessing the lowest eigenvalue, is initially selected at low pump powers. As the pump power increases, eigenstates with larger eigenvalues will be favored by the system. When the polariton-polariton interaction energy is shifted above that of the $K$-point modes, the system's emission becomes dominated by the degenerate states at the Brillouin zone corners. This mechanism is analogous to the blueshift behavior introduced by the pump power that is commonly observed in polariton condensates.


\bibliography{Reference}

\end{document}